\documentclass{aastex}
\usepackage{spr-astr-addons}
\usepackage{url}

\RequirePackage{color}
\begin{document}

\title{Radio Afterglow Rebrightening: Evidence for Multiple Active Phases in Gamma-Ray Burst Central Engines}
\slugcomment{}
%% Running heads
\shorttitle{Radio Afterglow Rebrightening: Evidence for Multiple Active Phases in Gamma-Ray Burst Central Engines}
\shortauthors{Li, Zhang \& Rice}

\author{Long-Biao Li\altaffilmark{1}}
\altaffiltext{1}{Guizhou University, Department of Physics, College of Sciences, Guiyang 550025, China}
\and
\author{Zhi-Bin Zhang\altaffilmark{1,2,*}}
\altaffiltext{1}{Guizhou University, Department of Physics, College of Sciences, Guiyang 550025, China}
\altaffiltext{2}{Department of Physics and Astronomy, University of Nevada, Las Vegas, NV 89154, USA}
\altaffiltext{*}{E-mail: sci.zbzhang@gzu.edu.cn}
\author{Jared Rice\altaffilmark{2}}
\altaffiltext{2}{Department of Physics and Astronomy, University of Nevada, Las Vegas, NV 89154, USA}

%\onecolumn
\begin{abstract}
  The rebrightening phenomenon is an interesting feature in some X-ray, optical, and radio afterglows of
  gamma-ray bursts (GRBs). Here, we propose a possible energy-supply assumption to explain the
  rebrightenings of radio afterglows, in which the central engine with multiple active phases
  can supply at least two GRB pulses in a typical GRB duration time. Considering the case of double
  pulses supplied by the central engine, the double pulses have separate physical parameters,
  except for the number density of the surrounding interstellar medium (ISM).
  Their independent radio afterglows are integrated by the ground detectors to form the rebrightening phenomenon.
  In this Letter, we firstly simulate diverse rebrightening light curves under consideration of different
  and independent physical parameters. Using this assumption, we also give our best fit to the radio afterglow of
  GRB 970508 at three frequencies of 1.43, 4.86, and 8.46 GHz. We suggest that the central engine may be
  active continuously at a timescale longer than that of a typical GRB duration time as many authors have suggested,
  and that it may supply enough energy to cause the long-lasting rebrightenings observed in some GRB afterglows.
\end{abstract}


%\onecolumn

%%%%%%%%%%%%%%%%%%%%%%%%%%%%%%%%%%%%%%%%%%%%%%%%%%%%%%%%%%%%
\section{Introduction}

Gamma-ray bursts (GRBs) are believed to be the brightest electromagnetic events in the universe.
These transient events in gamma-rays are usually followed by long-lived afterglows in X-ray,
optical, and radio bands. They were poorly understood until Feb 28, 1997 when the first afterglow
was detected \citep{Groot1998}, and when GRB 970508 was the first burst with an observed radio afterglow \citep{Frail1997}.
The fireball-shock model with synchrotron emission coming from the forward shock of ejecta plowing
into an external medium is successful in explaining the main features of GRB afterglows
\citep[see][for reviews]{Meszaros2002, Piran1999, Piran2005, Rees1994, Zhang2007, Zhang2004}.
Within this framework, many theoretical models have been proposed to explain the physical nature of GRBs
and their afterglows. GRBs are classified into two types, namely long GRBs and short GRBs,
according to the burst durations from CGRO/BATSE \citep{Kouveliotou1993}, Swift/BAT \citep{Zhang2008} and
Fermi/GBM \citep{Tarnopolski2015a, Tarnopolski2015b} in both observer and rest frames.
Generally, long GRBs should originate from the collapse of massive stars \citep{Woosley1993, MacFadyen1999},
and short GRBs could be connected with the coalescence of two compact objects \citep{Narayan1992, Gehrels2005, Nakar2007}.
Interestingly, the two types of bursts hold the consistent energy correlation
between peak energy and isotropic luminosity, namely $L_{iso} \sim  E_{p}^{1.7}$, which
implies an origin of thermal mechanism instead of single synchrotron radiation for
the GRB promopt emissions \citep{Zhang2012}.

As more afterglows have been observed, especially after the launch of Swift satellite in 2004,
studying afterglows has come into a new full-bandwidth era. In the meantime, many unexpected
and unusual phenomena, such as rebrightenings in multiple bands, have been observed in
afterglows \citep[see][for a review]{Zhang2007}. Rebrightening is an interesting
behavior among some GRBs. It was first discovered in the X-ray and optical afterglows of GRB 970508
\citep{Piro1998, Galama1998}. It was also discussed by \citet{Deng2010} in radio bands recently.
These rebrightenings are difficult to explain by the standard fireball-shock model,
since they exhibit a more complex decay other than a simple power-law.
As discussed by some authors \citep{Geng2013, Yu2013, Yu2015}, several interpretations
had been proposed to figure out the rebrightenings, e.g. the density jump model \citep{Dai2003, Tam2005},
the two-component jet model \citep{Huang2004, Liu2008}, the energy injection model
\citep{Dai1998, Huang2006, Geng2013, Yu2013, Yu2015} and the microphysics variation mechanism \citep{Kong2010}.

The plateau and flares in the X-ray afterglows indicate that the central engine
of GRBs should be active much longer than the prompt emission in gamma-rays
\citep{van Paradijs2000, Zhang2007, Lazzati2007, Geng2013}.
It had been pointed out that GRB pulses could reproduce the temporal activity of the inner engine
\citep{Kobayashi1997, Daigne1998, Nakar2002, Zhang2007a, Zhang2007b}.
\citet{Dermer2004} found that GRB pulses were useful for interpreting whether their
sources require central engines to be long-lasting or short-lived. In addition, the ultra-long duration
observed with multiple peaks would be related with the central engine activity \citep{ZhangBB2014}.
\citet{Laskar2015} modelled the re-brightening episodes with energy injection
into the forward blastwave, and considered that the phenomenon of energy injection is
ubiquitous in long GRBs, with rebrightening episodes likely due to extreme injection events.

Many rebrightenings in X-ray and optical afterglows have been studied, and
radio observations are vital for constraining the physical parameters. Therefore, in this letter,
we focus on rebrightenings in the radio band. We assume that the central engine can
produce at least two pulses in a typical GRB duration time. Considering different sets
of physical parameters, we simulate four rebrightening instances of radio afterglows.
Finally, we use our model to describe a specific burst, GRB 970508. The dynamical model
of GRB afterglows and our energy-supply assumption are introduced in Section 2.
In Section 3, we present our numerical results. Our conclusions are given in Section 4.

\begin{figure*}
  \centering
  \includegraphics[width=16cm]{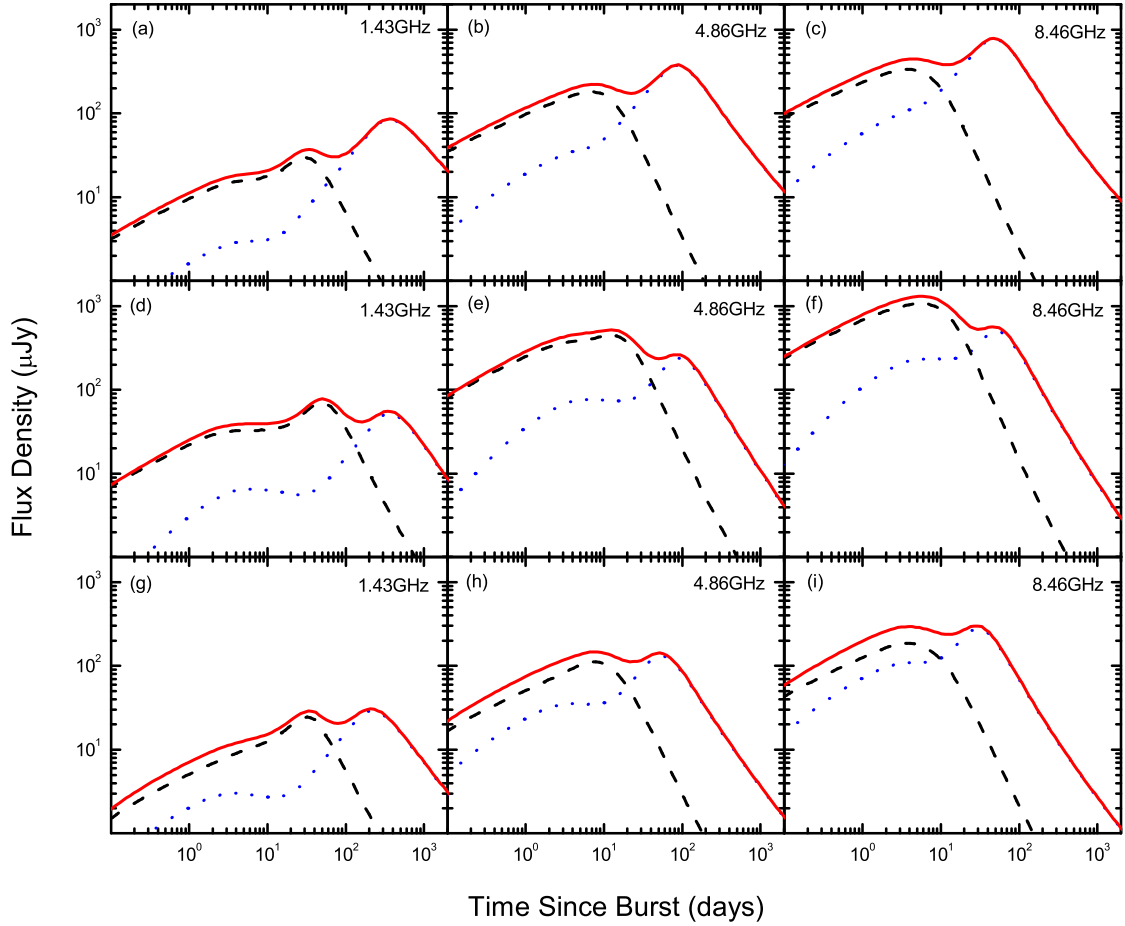}
  \caption{Three rebrightening instances of radio afterglow light curves at a redshift $z=1.0$.
           Every instance is displayed at three radio bands 1.43, 4.86, and 8.46 GHz.
           In each panel, radio afterglow light curves of the total burst, first, and second pulse
           are marked by solid, dotted, and dash lines, respectively.
           Panels (a) - (c) correspond to the first instance in Table 1.
           Similarly, panels (d) - (f) correspond to the second instance, and Panels (g) - (i)
           correspond to the third instance. Detailed parameter values used in the calculations
           are given in Table 1.}\label{Fig1}
\end{figure*}

\section{Model}

Based on the standard fireball-shock model, \citet{Huang1998, Huang1999a, Huang1999b, Huang2000a, Huang2000b}
proposed a dynamical model describing the evolution of external shocks and GRB afterglows.
The dynamical model is characterized by a system of four differential equations and is valid in both
the ultra-relativistic and non-relativistic (Newtonian) shock dynamical phase. Meanwhile,
this model takes the equal arrive time surface, the electron cooling, and the lateral expansion
into account. Many authors have applied this afterglow model to interpret a number of multiple-band observations
\citep[e.g.][]{Dai2005, Huang2000c, Kong2009, Wei2002, Wu2004, Xu2010, Geng2013, Yu2013, Yu2015}.

In fact, both the collapse of massive stars and the coalescence of two compact objects
are continuous activities, and may be accompanied by sustained expansion and contraction.
According to \citet{van Paradijs2000}, \citet{Zhang2007}, \citet{Lazzati2007} and \citet{Geng2013},
the central engines that provide enough energy to produce GRBs may be active continuously at a
timescale longer than that of a typical prompt gamma-ray duration time. Hence, considering
the observed multiple peaks of prompt emission, we assume that a central engine with multiple
active phases can supply multiple GRB pulses in a typical GRB duration time.

Here we assume that there are two pulses supplied by the central engine,
and define the double-pulse to be the first and second pulse respectively throughout
this letter. These two pluses constitute the total burst in our observations.
In addition, we also assume that there is no direct connection between the two pulses.
They have separate physical parameters, except for the number density of the surrounding
interstellar medium (ISM).
According to \citet{Chandra2012}, the radio afterglows which have been observed tend to occur
in a narrow range of surrounding ISM number density. At higher ISM number density, synchrotron
self-absorption effects suppress the radio afterglow strength for a long time.
This may imply that radio afterglows need long-time observations in order to be detected.
Hence we assume that for the two-step case, when the first pulse sweeps up the surrounding ISM,
the corresponding external shock will absorb a fraction of the ISM, which attenuates the number density of
the ISM surrounding the source. This means that the light curves are likely to show two peaks,
which is called the rebrightening phenomenon.

\section{Numerical Results}

\begin{table}[t]
  %\scriptsize
  \footnotesize
  \centering
  %\begin{minipage}{120mm}
  \caption{Physical Parameters of Particular Double-Pulses.}
  \scalebox{0.8}{
  \begin{tabular}{lccccccc}
  \hline
  \noalign{\smallskip}
   GRB Source &  $E_{iso}$  & $\gamma_{0}$ & $\theta_{0}$ & n & p & $\xi_{e}$ & $\xi^2_{B}$  \\

    & ($10^{52}$ergs) &  & (rad) & ($cm^{-3}$) & & &   \\
  \noalign{\smallskip}
  \hline
  \noalign{\smallskip}
                  \multicolumn{8}{c}{The First Rebrightening Instance} \\
  \noalign{\smallskip}
  \hline
  \noalign{\smallskip}
  First pulse         & $6.0$  & $300$  & 0.15 & 65.0 & 1.9 & 0.20  & 0.15   \\
  Second pulse        & $0.5$  & $100$  & 0.15 & 1.0  & 2.1 & 0.10  & 0.05   \\
  \hline
  \noalign{\smallskip}
  \multicolumn{8}{c}{The Second Rebrightening Instance} \\
  \noalign{\smallskip}
  \hline
  \noalign{\smallskip}
  First pulse         & $2.3$ & $300$ & 0.13  & 63.0 & 2.2 & 0.20   & 0.15    \\
  Second pulse        & $4.8$ & $100$ & 0.08  & 1.0  & 2.1 & 0.10   & 0.05    \\
  \hline
  \noalign{\smallskip}
  \multicolumn{8}{c}{The Third Rebrightening Instance} \\
  \noalign{\smallskip}
  \hline
  \noalign{\smallskip}
  First pulse         & $1.1$  & $300$ & 0.15  & 65.0 & 2.1 & 0.20   & 0.10     \\
  Second pulse        & $0.5$  & $30$  & 0.15  & 1.4  & 2.1 & 0.10   & 0.05    \\
  %\hline
  %\noalign{\smallskip}
  %\multicolumn{8}{c}{The Fourth Rebrightening Instance} \\
  %\noalign{\smallskip}
  %\hline
  %\noalign{\smallskip}
  %First pulse         & $3.0$  & $300$  & 0.18 & 35.0 & 1.9 & 0.05  & 0.05     \\
  %Second pulse        & $0.4$  & $200$  & 0.18 & 1.0  & 2.1 & 0.10  & 0.002     \\
  \hline
  \noalign{\smallskip}
  \multicolumn{8}{c}{The Radio Rebrightening of GRB 970508} \\
  \noalign{\smallskip}
  \hline
  \noalign{\smallskip}
  First pulse      & $6.0$  & $300$ & 0.10  & 82.3 & 2.1 & 0.13  & 0.24    \\
  Second pulse     & $8.0$  & $100$ & 0.07  & 2.5  & 1.9 & 0.11  & 0.15    \\
  \hline
  \end{tabular}
  }
  %\end{minipage}
 \label{Tab1}
\end{table}

For the purpose of simplicity, two pulses are assumed to occur independently for a given GRB source.
We calculate the radio afterglow light curves of the two pulses separately and add their emissions to
get the total afterglow fluxes from the jetted outflows.
Note that all physical parameters of the two pulses used in the simulations are listed in Table \ref{Tab1},
where $E_{iso}$ is the initial isotropic energy, $\gamma_{0}$ is the initial bulk Lorentz factor,
$\theta_{0}$ is the initial half-opening angle of the jet, $n$ is the number density of ISM,
$p$ is the electron distribution index,
$\xi_{e}$ and $\xi^2_{B}$ are respectively electron energy fraction and magnetic energy fraction.
For simplicity, the radiative efficiency is taken to be $\epsilon=0$ for a completely
adiabatic case. We also assume that the viewing angle between the axis of jet and the line
of sight $\theta_{obs}$ is zero.
In the meantime, we take a redshift $z=1.0$. % and cosmological constants with
%$\Omega_{M}=0.30, \Omega_{\Lambda}=0.70$ and $H_{0}=71\,km\,s^{-1}\,Mpc^{-1}$.
Note that the number density of the ISM corresponding to the second pulse is about
one order of magnitude lower than that of the first one, as shown in Table \ref{Tab1}.

Figure \ref{Fig1} illustrates our numerical results at the observed radio frequencies
1.43, 4.86, and 8.46 GHz by applying the model presented in Section 2 and using the
physical parameters of the corresponding rebrightening instance in Table \ref{Tab1}.
In each panel of Fig. \ref{Fig1}, there are three light curves, corresponding to
the radio afterglows of the first pulse, second pulse, and the total burst, respectively.

Fig. \ref{Fig1} (a) - (c) panels show the first rebrightening instance; the total
light curves show two peaks, with the former peak lower than the latter one. At 8.46 GHz,
the time of these two peaks are $\sim$ 4.5 days and 47.3 days after burst, respectively.
The corresponding flux densities are 445 and 787 $\mu$Jy. For the first instance, an obvious
rebrightening phenomenon can be observed at 8.46 GHz.
Note that the isotropic energy of the first pulse is larger than that of the second one,
so that the peak of the first burst is higher than that of the second one, which is more
obvious at higher frequencies. It is because the radio afterglow strength strongly
depends upon the kinetic energy of the burst \citep{Chandra2012}.

Fig. \ref{Fig1} (d) - (f) panels show a different rebrightening, which is that the
former peak is higher than the latter one. For the corresponding second instance in Table \ref{Tab1},
the isotropic energy of the first burst is smaller than that of the second burst.
At 8.46 GHz, the two peak flux densities of the total burst are 1305 and 266 $\mu$Jy
at $\sim$ 6 days and 45 days after burst, respectively.

The third instance is displayed in Fig. \ref{Fig1} (g) - (i) panels at the three corresponding radio frequencies.
In each panel, the two peaks of the total light curves have a similar height.
For the observational frequency $\nu=1.43$ GHz, both peak flux densities are $\sim$ 30 $\mu$Jy
at 33.2 days and 208.4 days after burst, respectively.

\begin{figure*}
  \centering
  \includegraphics[width=17.3cm]{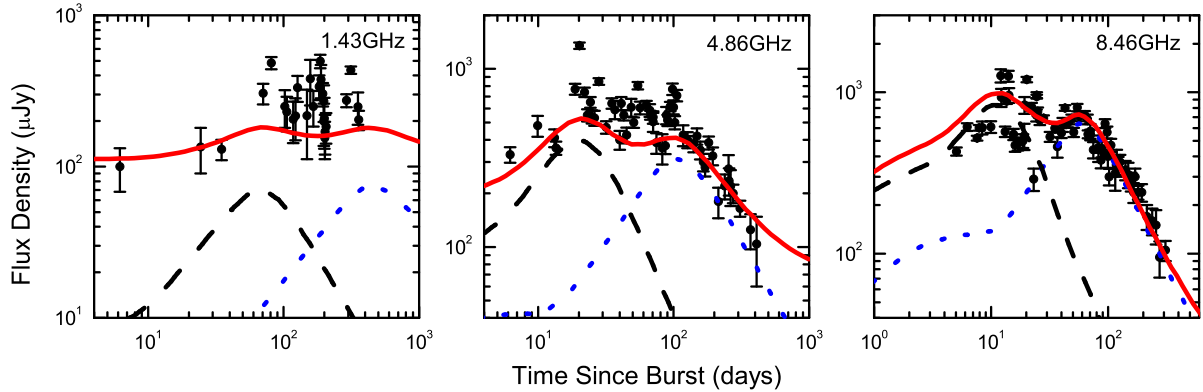}
  \caption{Comparing our model to the radio afterglow of GRB 970508 at 1.43, 4.86, and 8.46 GHz. Filled circles with error bars are
           the observational data referred in \citet{Frail2000}.  The dotted and dash lines stand for the first and
           second pulse afterglow light curves, correspondingly. The solid line is the expected total afterglow light
           curve, which is the sum of the calculated double-pulse afterglow light curves, after taking the flux
           densities of the host galaxy into account.
           }\label{Fig2}

\end{figure*}

Meanwhile, we use the energy-supply model to recalculate the radio afterglow light curve of
GRB 970508 at 1.43, 4.86, and 8.46 GHz. GRB 970508 was the first burst with an observed radio afterglow \citep{Frail1997}.
\citet{Panaitescu1998} investigated the rebrightening of the GRB 970508 afterglow in X-ray and optical bands,
and suggested that the rebrightening may be due to energy injection by a long-lived central engine.
We compare the radio afterglows of GRB 970508 \citep{Frail2000} with our energy-supply model in Fig. \ref{Fig2}.
Note that the contribution from the host has been taken into account, and we assume that the host fluxes
are respectively 120, 50, and 8 $\mu$Jy at 1.43, 4.86, and 8.46 GHz.
In the third panel of Fig. \ref{Fig2}, we find that the expected total afterglow light curve is relatively
consistent with the observational data, especially in later times. However, our model does not match
the data very well at an early stage of 30 days after the burst because of a larger fluctuation in observations.
This rapid variation is usually thought to be caused by interstellar scintillation when the source has
an apparent diameter of less than 3 micro-arcseconds \citep{Schilling2002}, and should be more obvious
at lower frequency. It is valuable to point out that the radio afterglow of GRB 980703, similar to GRB 970508,
have also suffered from the scattering effect of interstellar plasma in our Galaxy \citep{Kong2009}.

\section{Conclusion}

Rapid rebrightening is an unusual phenomenon in multi-wavelength GRB afterglows. In this letter,
based on the afterglow dynamical model \citep{Huang1998, Huang1999a, Huang1999b, Huang2000a, Huang2000b},
we propose a possible energy-supply mechanism to explain the rebrightening of radio afterglows.
Generally, the central engine is considered to be active for a longer time than the
duration of the gamma-ray prompt emission \citep{van Paradijs2000, Zhang2007, Lazzati2007, Geng2013}.
Considering the multiple peaks shown in gamma-ray band observations, the central engine may have
multiple active phases and supply two or more GRB pulses in a short time interval. In the case of a
double-pulse, the first and second pulses are considered to have little influence on each other.
These two pulses are assumed to have separate physical parameters except for the number density of
the surrounding ISM. However, the two jets launched from the same central engine may have some
physical connection, for example, the first jet with larger Lorentz factor would have a narrower
opening angle while the second jet could have a wider opening angle.
We argue that the number density of the ISM surrounding the second pulse is thinner than that
surrounding the first one. We present three different rebrightening instances and redescribe
the light curves of the GRB 970508 radio afterglow at 1.43, 4.86, and 8.46 GHz,
which shows a rebrightening behavior.
For the above three rebrightening instances, their flux densities are in the observational range
of China's Five-hundred-meter Aperture Spherical radio Telescope (FAST), which is expected to be
completed in Sep. 2016 and will be the largest radio telescope in the world.
However, observational constraints of radio afterglows are relatively rare.
We expect that there will be more rebrightening phenomena observed by FAST.

\acknowledgments
We thank the referee for constructive suggestions that improve our paper greatly. We appreciate
Prof. Bing Zhang for his very helpful comments on this model. This work is partly supported by
the National Natural Science Foundation of China (Grant Nos. 11263002, 11311140248 and U1431126)
and Provincial Research Foundations (20134021, 201519).

\end{document}